\documentclass[
 amsmath,
 amssymb,
 reprint,
 superscriptaddress
]{revtex4-1}
\usepackage{graphicx}
\usepackage{dcolumn}
\usepackage{bm}

\usepackage[utf8]{inputenc}
\usepackage[T1]{fontenc}
\usepackage{mathptmx}
\usepackage{etoolbox}
\usepackage{amsmath}
\usepackage[margin=1in]{geometry}
\makeatletter
\def\@email#1#2{%
 \endgroup
 \patchcmd{\titleblock@produce}
  {\frontmatter@RRAPformat}
  {\frontmatter@RRAPformat{\produce@RRAP{*#1\href{mailto:#2}{#2}}}\frontmatter@RRAPformat}
  {}{}
}%
\makeatother
\begin{document}


\title[Sample title]{Sub-nanometer resolution of the nitrogen-vacancy center by
Fourier magnetic imaging}
\affiliation{Laboratory of Spin Magnetic Resonance, School of Physical Sciences, \\Anhui Province Key
Laboratory of Scientific Instrument Development and Application, \\University of Science and
Technology of China, Hefei 230026, China.}
\affiliation{Hefei National Research Center for Physical Sciences at the Microscale, \\University of Science
and Technology of China, Hefei 230026, China.}
\affiliation{Hefei National Laboratory, University of Science and Technology of China, Hefei 230088, China.}
\affiliation{School of Biomedical Engineering and Suzhou Institute for Advanced Research, \\University of
Science and Technology of China, Suzhou 215123, China}
\author{Peihan Lei}
\affiliation{Laboratory of Spin Magnetic Resonance, School of Physical Sciences, \\Anhui Province Key
Laboratory of Scientific Instrument Development and Application, \\University of Science and
Technology of China, Hefei 230026, China.}
\affiliation{Hefei National Laboratory, University of Science and Technology of China, Hefei 230088, China.}
\author{You Huang}
\affiliation{Laboratory of Spin Magnetic Resonance, School of Physical Sciences, \\Anhui Province Key
Laboratory of Scientific Instrument Development and Application, \\University of Science and
Technology of China, Hefei 230026, China.}
\author{Zhi Cheng}
\affiliation{Laboratory of Spin Magnetic Resonance, School of Physical Sciences, \\Anhui Province Key
Laboratory of Scientific Instrument Development and Application, \\University of Science and
Technology of China, Hefei 230026, China.}
\author{Fazhan Shi}
\affiliation{Laboratory of Spin Magnetic Resonance, School of Physical Sciences, \\Anhui Province Key
Laboratory of Scientific Instrument Development and Application, \\University of Science and
Technology of China, Hefei 230026, China.}
\affiliation{Hefei National Research Center for Physical Sciences at the Microscale, \\University of Science
and Technology of China, Hefei 230026, China.}
\affiliation{Hefei National Laboratory, University of Science and Technology of China, Hefei 230088, China.}
\affiliation{School of Biomedical Engineering and Suzhou Institute for Advanced Research, \\University of
Science and Technology of China, Suzhou 215123, China}
\author{Pengfei Wang}
 \altaffiliation{Corresponding author}
\email{wpf@ustc.edu.cn}
\affiliation{Laboratory of Spin Magnetic Resonance, School of Physical Sciences, \\Anhui Province Key
Laboratory of Scientific Instrument Development and Application, \\University of Science and
Technology of China, Hefei 230026, China.}
\affiliation{Hefei National Research Center for Physical Sciences at the Microscale, \\University of Science
and Technology of China, Hefei 230026, China.}
\affiliation{Hefei National Laboratory, University of Science and Technology of China, Hefei 230088, China.}



\date{}

\begin{abstract}
Solid-state spins in diamond are promising building blocks for quantum computing and quantum
sensing, both of which require precise nanoscale addressing of individual spins. To explore the
resolution limit of this approach, we demonstrate Fourier magnetic imaging of nitrogen-vacancy
centers in diamond under state-of-the-art conditions. We constructed a highly compact
experimental platform featuring thermal drift compensation under ambient conditions and
generated a pulsed magnetic field gradient of up to 13.5 G/\textmu m. By implementing the Fourier
magnetic imaging protocol, we achieved localization of a single nitrogen-vacancy center with a
spatial resolution of 0.28 ± 0.10 nm and a magnetic field measurement deviation of 9 nT. This
technique holds potential for applications such as localizing spins within proteins and cells. 
\end{abstract}

\maketitle


The nitrogen vacancy (NV) center in diamond is an atomic scale sensor with promised high spatial
resolution and high sensitivity under ambient condition owing to its long spin coherent time\cite{RN1,RN2,RN3},
efficient readout and control fidelity\cite{RN4,RN5,RN6} as well as the quantum scalability to multi-qubit system\cite{RN7,RN8,RN9,RN10}.
Based on these extraordinary properties, NV center has become one of the most promising candidates
for quantum information processors\cite{RN11,RN12,RN13,RN14} and highly sensitive quantum sensors with nanoscale
resolution\cite{RN15,RN16,RN17}. Over the past decades, quantum technology based on NV centers, has achieved
tremendous progress on quantum sensing\cite{RN18,RN19}, spintronics\cite{RN20,RN21}, advanced material\cite{RN22,RN23} and life
sciences\cite{RN24}.

In these applications, most of them rely on the spatial imaging of the electron spin in the NV center.
Currently, various methods exist for locating the NV center. For instance, confocal microscopy, as a
form of optical microscopy, achieves a spatial resolution of about 200 nm, while the super-resolution
STED improves this to well below 10 nm\cite{RN25}. Beyond optical approaches, the localization of NV
centers can be spatially encoded into the sub-nanometer level by a strong magnetic field gradient
generated by a scanning magnetic tip\cite{RN26}. More robustly, it can be encoded by the pulsed magnetic field
gradient from the current in the microwire and a Fourier magnetic resonance imaging scheme can be
adopted to nanometer resolution with high stability, simplified experimental setup and the multiplex
advantage into the parallel imaging\cite{RN27}. However, until now the highest spatial resolution or pixel
resolution is about 3.5 nm\cite{RN28}, far below the theoretical prediction. The decoherence time of NV center
and the noise under the high current pulse become the main limitation.

In this letter, we explore the realization of sub-nanometer resolution of NV center by combining the
stabilization improved experimental platform and magnetic Fourier imaging. Firstly, the stabilization
and thermal drift of optically detected magnetic resonance (ODMR) spectrometer is improved by
fiber-collected fluorescence collection and compact design. Secondly, a microwire is fabricated directly
on the diamond surface to generate the strong magnetic field gradient pulse. Finally, we demonstrate
the resolution of a single NV center of 0.28 nm in one dimension.

\begin{figure*}
\includegraphics{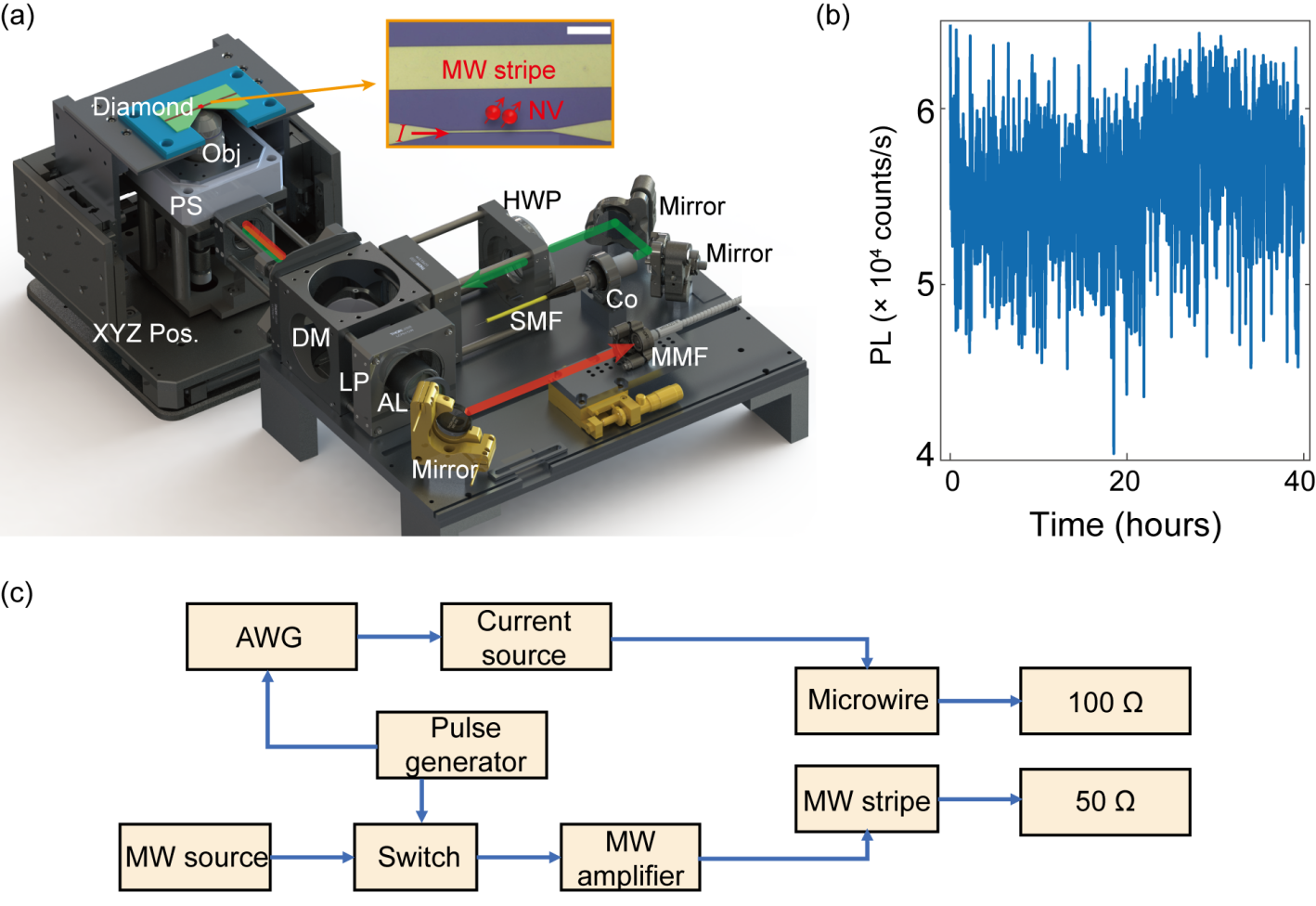}
\caption{\label{fig:1} The experimental setup. (a) The confocal microscope part of optically detected magnetic resonance (ODMR) platform. The overall outer dimension is 0.5(L) × 0.4(W) × 0.2 (H) $\mathrm{m}^{3}$. Inset: the microscopy photograph of the microwave (MW) stripe and gradient microwire fabricated on the diamond surface. The scale bar is 20 $\mu$m. Abbreviations: Obj: objective lens; XYZ Pos.: $xyz$ positioner; PS: $xyz$ piezo scanner; DM: dichroic mirror; LP: 650 nm long pass filter; AL: achromatic lens; SMF: single-mode fiber; HWP: half wave plate; Co: fiber collimator; MMF: multi-mode fiber with 50 $\mu$m core diameter. The SMF is connected to the laser pulse module and MMF is connected to APD. (b) The fluorescence tracking of a single NV center in 40 hours. The long time fluctuation is about ±10\%. The fast noise is due to the short sampling time of about 10 ms at each data point. (c) Electric circuits
for the generation of magnetic field gradient (MFG) pulse and
the synchronization
between MW and MFG.}
\end{figure*}

During the long time sampling in Fourier magnetic imaging, the drift of the platform will cause the
line broadening of the Fourier transformed positioning peak. To avoid the line broadening effect, we build a thermal drift compensated ODMR platform(Fig.~\ref{fig:1})
.
A home-built confocal microscope with
fiber-connected laser excitation and fluorescence collection is used to illuminate the nitrogen vacancy centers and readout the quantum state. We use 532 nm laser for excitation and collect the fluorescence in 650-800 nm. The single-mode fiber is used for collecting the modulated laser pulse from double-pass acoustic modulator module, and then guiding it to the dichroic mirror. At the scanning probe head, the laser is focused by a high-NA air objective lens and aligned with the NV center by a miniature piezo XYZ nano-scanner. The scanning probe head is also designed symmetrically and made of stainless steel to minimize the thermal expansion. The fluorescence from the NV center is collected by the objective lens, then focused into a multi-mode fiber with 50 \textmu m diameter and finally measured by fiber-based avalanche single photon counter module (SPCM). The alignment of focused laser and the fluorescence collection with the fiber is adjusted by tuning the rotation angle of the piezo controlled reflection mirrors, which can minimize the length of the optical path to $\sim$ 0.5 m. The environment is controlled with a temperature deviation of ±1 K.

We then investigate the long-time mechanical stability performance of the platform. The
temperature fluctuation is the dominant source of the thermal drift. Here, by tracking a single NV
center in every 2 minutes, we record the fluorescent photon count. During this measurement period, a
temperature sensor detected the environment temperature accordingly. By simultaneously recording the
temperature, the position of a single NV center, and the PL count, we studied the robustness of the
experimental setup about 40 hours [Fig.~\ref{fig:1}(b)\&(c)]. The temperature fluctuation is ±0.25 K, while the
drifts in the fluorescence is <10\%. The stability of the platform provide the base for long time high
resolution magnetic imaging.

To generate the strong magnetic field gradient, we fabricate a microwire directly on the diamond
surface, with the estimated current carrying ability above 0.1 A. The MW and MFG wire connections
are shown in Fig.~\ref{fig:1}(d). The MW and the current sent through the electrodes on the surface of the
diamond are triggered and synchronized by transistor-transistor logic (TTL) signals from the pulse
generator. Importantly, the synchronization between the MW and MFG pulses must be perfectly
aligned. Terminal resistances of 50 $\Omega$ and 100 $\Omega$ are applied at the end of the MW and current circuits,
respectively. The 50 $\Omega$ resistance serves as the impedance in the MW circuit, while the 100 $\Omega$
resistance acts as a voltage divider to protect the MFG electrode.

\begin{figure}
\includegraphics{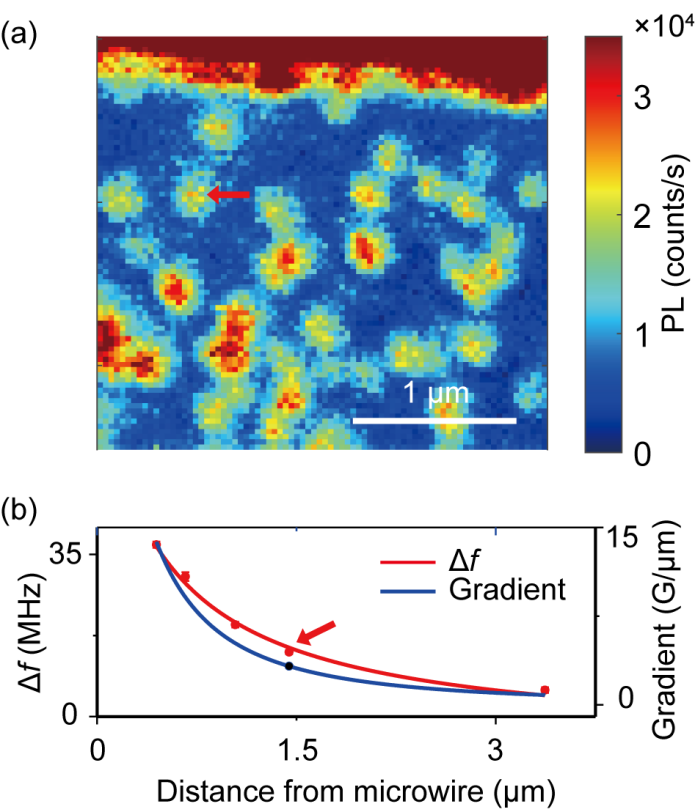}
\caption{\label{fig:2}  Characterization of the MFG. (a) the NV fluorescence map near the gradient microwire. On the top, the red
stripe is the gradient microwire. (b) $m_{S}=0$ to $m_{S}=+1$ transition frequency shift $\Delta f$ of several NV centers vs.
Positions (red dot) under a direct current $I$ = 1 mA. The red and blue curves show the fitting to the frequency and
the calculation of the MFG projected on the NV-[111] axis, respectively. The red arrows in both (a) and (b)
indicate the NV center for demonstration of Fourier magnetic imaging.
}
\end{figure}

We then measure the magnetic field gradient by measure the magnetic resonance spectrum of
several NV centers at different positions. 
Fig.~\ref{fig:2}(b)
presents the measured magnetic field projected on
NV-[111] quantum axis (red curve) and the calculated calibrated MFG (blue curve), obtained by
differentiating the magnetic field with respect to the distance ($G\ =\ \partial B/\partial x$). The black dot marks the
specific NV center selected to demonstrate the performance of our robust spectrometer. The magnetic
field data is derived from frequency shifts $\Delta f$ of five different NV centers (the red dots) in the same
quantum axis orientation surrounding the microwire, using the relation $B = \Delta f/\gamma_{NV}$, where $\Delta f = f - f_0$.
Here, $f$ and $f_0$ represent the ODMR spectral peaks with and without current applied to the microwire,
respectively. The measured magnetic field gradient is up to 13.5 G/$\mu$m with $I$ = 10 mA. We select a
single NV center near the MFG microwire in the confocal map [black dot in Fig.~\ref{fig:2}(b)] with the
magnetic field gradient of $G$ = (3.26 ± 0.06) G/$\mu$m and a longest decoherence time of over 1.2 ms for
the demonstration.

\begin{figure}
\includegraphics{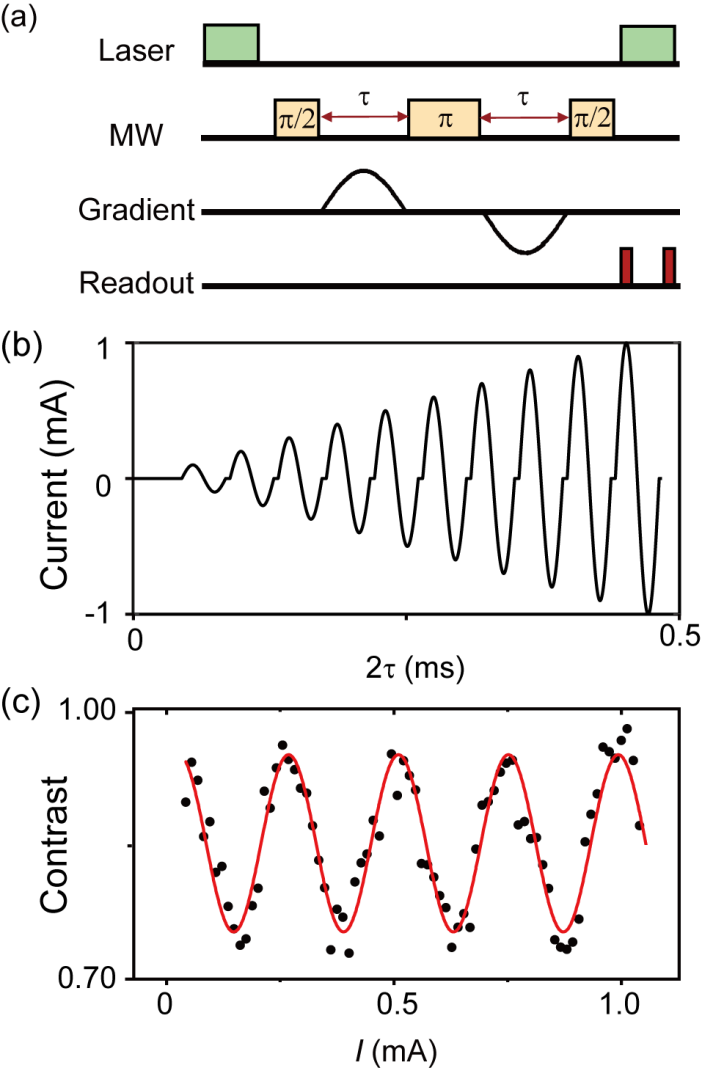}
\caption{\label{fig:3}  Architecture for nanoscale MFG technique. (a) The pulse sequence consisting of spin echo pulses and
synchronized MFG pulses for Fourier magnetic imaging. (b) The waveform for generating the MFG. It is not a
continuous sine curve, where the gap between two period is used for laser initialization and readout of NV center.
(c) The raw echo signal vs. current amplitude, with 2$\tau$=21 $\mu$s. The data is fitted by a cosine function.
}
\end{figure}

Fig.~\ref{fig:3}(a)
presents the pulse sequence for MFG imaging based on spin echo, where the MW and
MFG pulses are applied alternately through the microwire. Due to the intermediate $\pi$ pulse, only the
phase accumulated by the MFG is retained. Under the magnetic field gradient pulse, the spin
accumulates the relative quantum phase when preparing the spin into quantum superposition state in
advance. The detected signal (K-space signal) can then be expressed as a function of the MFG\cite{RN29}
\begin{equation}
    \begin{aligned}
        s\sim \mathrm{cos}\Big[ \int_{0}^{2\tau} G(x,t,I)x\mathrm{dt} \Big]=\mathrm{cos}(2\pi K\cdot x)
    \end{aligned}
\end{equation}
where $K = 2\gamma_{NV}\tau G/2\pi$ is the K-space parameter for Fourier imaging. Here, $\gamma_{NV} = 2\pi\ \times\ 2.8\ \mathrm{MHz/G}$
denotes the gyromagnetic ratio of the NV center, and 2$\tau$ represents the total evolution time of the pulse
sequence. $G$ and $x$ correspond to the MFG and the spatial position of the NV center, respectively. In the
experiment, the evolution time is fixed while the amplitude of the magnetic field gradient (MFG) is
linearly increased. As indicated by the above expression, a cosine signal is obtained when the MFG
amplitude increases linearly. It should be noted that such a signal can only be clearly resolved under
smoothly varying current pulse shape, such as sine shape, though a rectangular shape waveform can
maximum the quantum phase accumulation. This discrepancy can be attributed to two factors: first,
imperfect synchronization between the MFG and the microwave pulse disturbs the accumulated phase;
second, the magnetic field generated by the current in the gradient microwire compromises the
accuracy of the intermediate $\pi$ pulse, leading to distortion in the accumulated phase. Fig.~\ref{fig:3}(b) shows
the sine-wave mode current pulses and Fig.~\ref{fig:3}(c) shows the corresponding detected cosine signal with
the evolution time of $2\tau = 21\ \mu s$.

\begin{figure}
\includegraphics{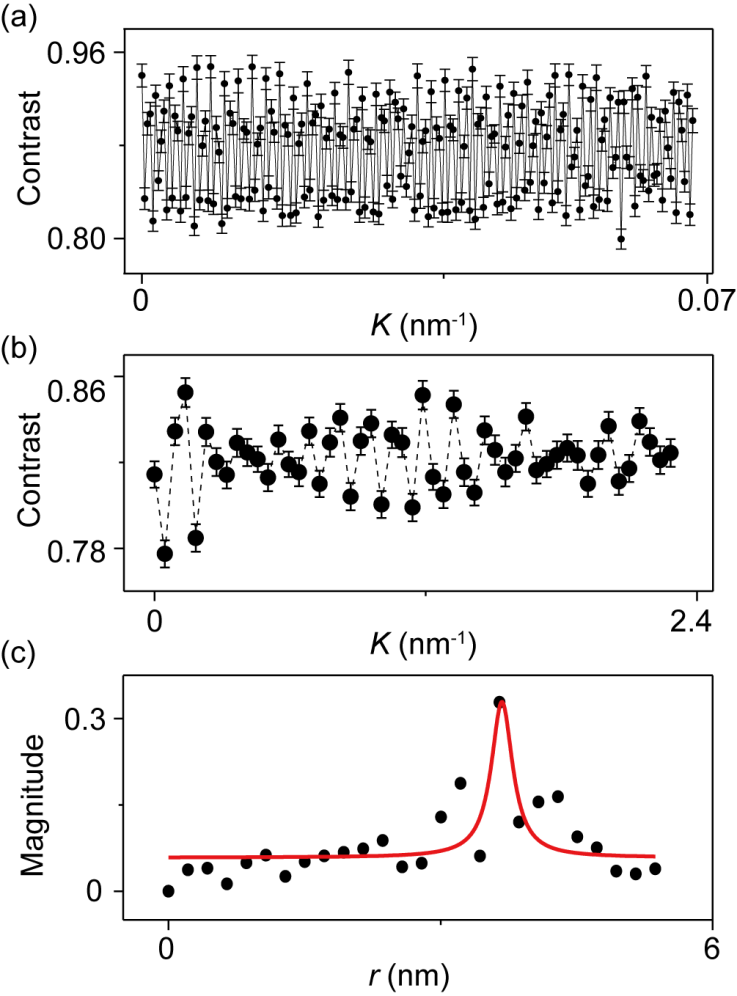}
\caption{\label{fig:4}  Demonstration of nanoscale localization of a single NV center by using the robust experimental apparatus.
(a) The K-space signal with $2\tau\ =\ 80\ \mu s$, from which the single frequency oscillation can be clearly seen. (b) The
detected signal in K-space, where the signal is obtained through undersampling as it consists of thousands of
oscillations in total. Here the maximum current sent through the microwire is $I_{MAX}=10\ \mathrm{mA}$ and the total evolution
time is $2\tau\ =\ 500\ \mu s$. (c) The real space localization of the single NV center after Fourier transformation of K-space
signal. The red solid curve is Lorentz fit and the corresponding full width at half maximum (FWHM) is 0.28 ± 0.10 nm
}
\end{figure}

We then perform the Fourier magnetic imaging sequence to explore the spatial resolution on the
platform. 
Fig.~\ref{fig:4}(a)
shows an example of the experimental result as a function of the K-space
parameter under $I_{MAX}=1\ \mathrm{mA}$ and $2\tau\ =\ 80\ \mu s$. Note that the total number of oscillation periods reaches
inaccessible several thousand if we acquire the full K-space data. The spatial resolution is determined
by either the maximum applied MFG or the total evolution time $2\tau$, as indicated by the K-space
parameter equation (1). In the next experiment with the longer evolution time of $2\tau\ =\ 500\ \mu s$ and $I_{MAX}=10\ \mathrm{mA}$, we selectively acquire the K-space, resulting in an undersampling original K-space signal in
Fig.~\ref{fig:4}(b), with the prior knowledge of a single NV center in the field of view. Finally, we obtain a
K-space result with K from 0 to 2.2834 $\mathrm{nm}^{-1}$, corresponding to a pixel resolution of ~ 0.22 nm. After
Fourier transformation of the full K-space signal yields the real-space localization of the NV center, as
shown in Fig.~\ref{fig:4}(c). The singular peak in real space underscores the exceptional stability of the
apparatus. The two smaller sideband peaks are likely attributable to current fluctuations through the
microwire, which also appears as the modulation in Fig.~\ref{fig:4}(b). By fitting the real-space data with a
Lorentzian function, a localization resolution of 0.28 ± 0.10 nm is achieved for a single NV center (Fig.~\ref{fig:4}(c)) and is close to the theoretical pixel resolution.

We also calculated the sensitivity of the NV center as a magnetic field sensor under this high
resolution as it is close to the theoretical angstrom scale. The sensitivity of the magnetic sensor can be
calculated by\cite{RN15,RN16,RN17}:
\begin{equation}
    \begin{aligned}
        \eta\ =\ |\mathrm{d}B/\mathrm{d}S|_{MAX}\sigma_{S}\notag
    \end{aligned}
\end{equation}
Where $|\mathrm{d}B/\mathrm{d}S|_{MAX}\ =\ (2\gamma\tau\alpha\beta)^{-1}$ is the maximum slope of the spin-echo signal variation with the
magnetic field amplitude, $\alpha$ is the contrast of the spin echo signal and $\beta$ is the average photon counts in
one experiment. $\sigma_{S}$ is the photon-shot-noise related standard deviation of the spin echo signal under
the sampling bandwidth of 1 Hz. In our experiment, $\alpha\ \sim\ 0.08$, $\beta\ \sim\ 0.02$ and $\sigma_{S}=0.06\ \mathrm{Hz}^{1/2}$. This gives
the magnetic sensitivity of about $\eta\sim0.2\ \mu\mathrm{T}\ \mathrm{Hz}^{1/2}$ and a magnetic field deviation of 9 nT after $10^{6}$ times
average.

It should be noted that the spatial resolution and the full width at half maximum (FWHM)
approaches the lattice size of the diamond but the shape of the electron wave function as the spin
carrier is not observed, whose radius is $R_{\mathrm{vac}}\approx 0.6\ \mathrm{nm}$\cite{RN30}. Under the ultra fast movement of the electrons
in electron cloud, the observing time, defined by the quantum evolution time, is much longer than the
moving period in electron cloud. It is inferred that the location of the NV center reveals the expected
average position of the spin, but not the transient position of the electron.

In summary, to investigate the state-of-the-art resolution in Fourier magnetic imaging, we have
designed and constructed a fiber-connected compact, robust and low-drift ODMR platform. We
demonstrated the one dimensional Fourier magnetic imaging of a single NV center with the resolution
of 0.28 nm. This Fourier imaging spectrometer with fiber-based collection is also suitable for
addressing and manipulating spins at the nanometer scale, potentially for multi-qubit quantum
processors. The achieved spatial scale is already close to the size of amino acids in protein molecules,
which also allows for the conformation analysis of single protein combing with single molecule
magnetic resonance technique\cite{RN31}.
\begin{acknowledgments}
This work was supported by the Chinese Academy of Sciences (Grants No.
ZDZBGCH2021002), the National Natural Science Foundation of China (Grants No. T2125011),
Quantum Science and Technology-National Science and Technology Major Project (Grants No.
2021ZD0303204), the instrument application demonstration center of Academy of Sciences, the
Fundamental Research Funds for the Central Universities and the USTC Tang scholar. This work was
partially carried out at the USTC Center for Micro and Nanoscale Research and Fabrication.
\end{acknowledgments}
\section*{Data Availability Statement}
The data that support the findings of this study are available from the corresponding author upon
reasonable request.

\nocite{*}
\bibliographystyle{unsrt}
\bibliography{ref1}

\begin{thebibliography}{10}

\bibitem{RN1}
G.~Balasubramanian, P.~Neumann, D.~Twitchen, M.~Markham, R.~Kolesov, N.~Mizuochi, J.~Isoya, J.~Achard, J.~Beck, J.~Tissler, V.~Jacques, P.~R. Hemmer, F.~Jelezko, and J.~Wrachtrup.
\newblock Ultralong spin coherence time in isotopically engineered diamond.
\newblock {\em Nat. Mater.}, 8(5):383--387, 2009.

\bibitem{RN2}
N.~Bar-Gill, L.~M. Pham, A.~Jarmola, D.~Budker, and R.~L. Walsworth.
\newblock Solid-state electronic spin coherence time approaching one second.
\newblock {\em Nat. Commun.}, 4, 2013.

\bibitem{RN3}
B.~Naydenov, F.~Reinhard, A.~Lämmle, V.~Richter, R.~Kalish, U.~F.~S. D'Haenens-Johansson, M.~Newton, F.~Jelezko, and J.~Wrachtrup.
\newblock Increasing the coherence time of single electron spins in diamond by high temperature annealing.
\newblock {\em Appl. Phys. Lett.}, 97(24), 2010.

\bibitem{RN4}
B.~J. Shields, Q.~P. Unterreithmeier, N.~P. de~Leon, H.~Park, and M.~D. Lukin.
\newblock Efficient readout of a single spin state in diamond via spin-to-charge conversion.
\newblock {\em Phys. Rev. Lett.}, 114(13), 2015.

\bibitem{RN5}
D.~M. Irber, F.~Poggiali, F.~Kong, M.~Kieschnick, T.~Lühmann, D.~Kwiatkowski, J.~Meijer, J.~F. Du, F.~Z. Shi, and F.~Reinhard.
\newblock Robust all-optical single-shot readout of nitrogen-vacancy centers in diamond.
\newblock {\em Nat. Commun.}, 12(1), 2021.

\bibitem{RN6}
Q.~Zhang, Y.~H. Guo, W.~T. Ji, M.~Q. Wang, J.~Yin, F.~Kong, Y.~H. Lin, C.~M. Yin, F.~Z. Shi, Y.~Wang, and J.~F. Du.
\newblock High-fidelity single-shot readout of single electron spin in diamond with spin-to-charge conversion.
\newblock {\em Nat. Commun.}, 12(1), 2021.

\bibitem{RN7}
L.~Childress and R.~Hanson.
\newblock Diamond nv centers for quantum computing and quantum networks.
\newblock {\em Mrs Bulletin}, 38(2):134--138, 2013.

\bibitem{RN8}
G.~Waldherr, Y.~Wang, S.~Zaiser, M.~Jamali, T.~Schulte-Herbrüggen, H.~Abe, T.~Ohshima, J.~Isoya, J.~F. Du, P.~Neumann, and J.~Wrachtrup.
\newblock Quantum error correction in a solid-state hybrid spin register.
\newblock {\em Nature}, 506(7487):204--207, 2014.

\bibitem{RN9}
S.~Pezzagna and J.~Meijer.
\newblock Quantum computer based on color centers in diamond.
\newblock {\em Appl. Phys. Rev.}, 8(8), 2021.

\bibitem{RN10}
C.~E. Bradley, J.~Randall, M.~H. Abobeih, R.~C. Berrevoets, M.~J. Degen, M.~A. Bakker, M.~Markham, D.~J. Twitchen, and T.~H. Taminiau.
\newblock A ten-qubit solid-state spin register with quantum memory up to one minute.
\newblock {\em Phys. Rev. X}, 9(3), 2019.

\bibitem{RN11}
J.~Wrachtrup and F.~Jelezko.
\newblock Processing quantum information in diamond.
\newblock {\em J. Phys.: Condens. Matter}, 18(21):S807--S824, 2006.

\bibitem{RN12}
H.~Bernien, B.~Hensen, W.~Pfaff, G.~Koolstra, M.~S. Blok, L.~Robledo, T.~H. Taminiau, M.~Markham, D.~J. Twitchen, L.~Childress, and R.~Hanson.
\newblock Heralded entanglement between solid-state qubits separated by three metres.
\newblock {\em Nature}, 497(7447):86--90, 2013.

\bibitem{RN13}
B.~Hensen, H.~Bernien, A.~E. Dréau, A.~Reiserer, N.~Kalb, M.~S. Blok, J.~Ruitenberg, R.~F.~L. Vermeulen, R.~N. Schouten, C.~Abellán, W.~Amaya, V.~Pruneri, M.~W. Mitchell, M.~Markham, D.~J. Twitchen, D.~Elkouss, S.~Wehner, T.~H. Taminiau, and R.~Hanson.
\newblock Loophole-free bell inequality violation using electron spins separated by 1.3 kilometres.
\newblock {\em Nature}, 526(7575):682--686, 2015.

\bibitem{RN14}
M.~H. Abobeih, Y.~Wang, J.~Randall, S.~J.~H. Loenen, C.~E. Bradley, M.~Markham, D.~J. Twitchen, B.~M. Terhal, and T.~H. Taminiau.
\newblock Fault-tolerant operation of a logical qubit in a diamond quantum processor.
\newblock {\em Nature}, 606(7916):884--+, 2022.

\bibitem{RN15}
J.~M. Taylor, P.~Cappellaro, L.~Childress, L.~Jiang, D.~Budker, P.~R. Hemmer, A.~Yacoby, R.~Walsworth, and M.~D. Lukin.
\newblock High-sensitivity diamond magnetometer with nanoscale resolution.
\newblock {\em Nat. Phys.}, 4(10):810--816, 2008.

\bibitem{RN16}
G.~Balasubramanian, I.~Y. Chan, R.~Kolesov, M.~Al-Hmoud, J.~Tisler, C.~Shin, C.~Kim, A.~Wojcik, P.~R. Hemmer, A.~Krueger, T.~Hanke, A.~Leitenstorfer, R.~Bratschitsch, F.~Jelezko, and J.~Wrachtrup.
\newblock Nanoscale imaging magnetometry with diamond spins under ambient conditions.
\newblock {\em Nature}, 455(7213):648--U46, 2008.

\bibitem{RN17}
C.~L. Degen, F.~Reinhard, and P.~Cappellaro.
\newblock Quantum sensing.
\newblock {\em Rev. Mod. Phys.}, 89(3), 2017.

\bibitem{RN18}
T.~Staudacher, F.~Shi, S.~Pezzagna, J.~Meijer, J.~Du, C.~A. Meriles, F.~Reinhard, and J.~Wrachtrup.
\newblock Nuclear magnetic resonance spectroscopy on a (5-nanometer) sample volume.
\newblock {\em Science}, 339(6119):561--563, 2013.

\bibitem{RN19}
H.~J. Mamin, M.~Kim, M.~H. Sherwood, C.~T. Rettner, K.~Ohno, D.~D. Awschalom, and D.~Rugar.
\newblock Nanoscale nuclear magnetic resonance with a nitrogen-vacancy spin sensor.
\newblock {\em Science}, 339(6119):557--560, 2013.

\bibitem{RN20}
F.~Jelezko and J.~Wrachtrup.
\newblock Focus on diamond-based photonics and spintronics.
\newblock {\em New J. Phys.}, 14, 2012.

\bibitem{RN21}
D.~Prananto, Y.~Kainuma, K.~Hayashi, N.~Mizuochi, K.~Uchida, and T.~An.
\newblock Probing thermal magnon current mediated by coherent magnon via nitrogen-vacancy centers in diamond.
\newblock {\em Phys. Rev. Appl.}, 16(6), 2021.

\bibitem{RN22}
I.~Lovchinsky, J.~D. Sanchez-Yamagishi, E.~K. Urbach, S.~Choi, S.~Fang, T.~I. Andersen, K.~Watanabe, T.~Taniguchi, A.~Bylinskii, E.~Kaxiras, P.~Kim, H.~Park, and M.~D. Lukin.
\newblock Magnetic resonance spectroscopy of an atomically thin material using a single-spin qubit.
\newblock {\em Science}, 355(6324):503--+, 2017.

\bibitem{RN23}
L.~Thiel, Z.~Wang, M.~A. Tschudin, D.~Rohner, I.~Gutiérrez-Lezama, N.~Ubrig, M.~Gibertini, E.~Giannini, A.~F. Morpurgo, and P.~Maletinsky.
\newblock Probing magnetism in 2d materials at the nanoscale with single-spin microscopy.
\newblock {\em Science}, 364(6444):973--+, 2019.

\bibitem{RN24}
N.~Aslam, H.~Y. Zhou, E.~K. Urbach, M.~J. Turner, R.~L. Walsworth, M.~D. Lukin, and H.~Park.
\newblock Quantum sensors for biomedical applications.
\newblock {\em Nat. Rev. Phys.}, 5(3):157--169, 2023.

\bibitem{RN25}
E.~Rittweger, K.~Y. Han, S.~E. Irvine, C.~Eggeling, and S.~W. Hell.
\newblock Sted microscopy reveals crystal colour centres with nanometric resolution.
\newblock {\em Nat. Photonics}, 3(3):144--147, 2009.

\bibitem{RN26}
S.~J. DeVience, L.~M. Pham, I.~Lovchinsky, A.~O. Sushkov, N.~Bar-Gill, C.~Belthangady, F.~Casola, M.~Corbett, H.~L. Zhang, M.~Lukin, H.~Park, A.~Yacoby, and R.~L. Walsworth.
\newblock Nanoscale nmr spectroscopy and imaging of multiple nuclear species.
\newblock {\em Nat. Nanotechnol.}, 10(2):129--134, 2015.

\bibitem{RN27}
Z.~Z. Guo, Y.~Huang, M.~C. Cai, C.~X. Li, M.~Z. Shen, M.~Q. Wang, P.~Yu, Y.~Wang, F.~Z. Shi, P.~F. Wang, and J.~F. Du.
\newblock Wide-field fourier magnetic imaging with electron spins in diamond.
\newblock {\em npj Quantum Inf.}, 10(1), 2024.

\bibitem{RN28}
K.~Arai, C.~Belthangady, H.~Zhang, N.~Bar-Gill, S.~J. DeVience, P.~Cappellaro, A.~Yacoby, and R.~L. Walsworth.
\newblock Fourier magnetic imaging with nanoscale resolution and compressed sensing speed-up using electronic spins in diamond.
\newblock {\em Nat. Nanotechnol.}, 10(10):859--864, 2015.

\bibitem{RN29}
Y.~Huang, M.~S. Guo, M.~Z. Shen, P.~Yu, M.~Q. Wang, Y.~Wang, C.~K. Duan, F.~Z. Shi, P.~F. Wang, and J.~F. Du.
\newblock Superresolution localization of nitrogen-vacancy centers in diamond with quantum-controlled photoswitching.
\newblock {\em Phys. Rev. A}, 102(4), 2020.

\bibitem{RN30}
A.~Gali, M.~Fyta, and E.~Kaxiras.
\newblock Ab initio supercell calculations on nitrogen-vacancy center in diamond: Electronic structure and hyperfine tensors.
\newblock {\em Phys. Rev. B}, 77(15), 2008.

\bibitem{RN31}
F.~Z. Shi, Q.~Zhang, P.~F. Wang, H.~B. Sun, J.~R. Wang, X.~Rong, M.~Chen, C.~Y. Ju, F.~Reinhard, H.~W. Chen, J.~Wrachtrup, J.~F. Wang, and J.~F. Du.
\newblock Single-protein spin resonance spectroscopy under ambient conditions.
\newblock {\em Science}, 347(6226):1135--1138, 2015.

\end{thebibliography}

\end{document}